\documentclass[prl,twocolumn]{revtex4}%
\usepackage{amsmath}
\usepackage{amsfonts}
\usepackage{amssymb}
\usepackage{graphicx}
\begin{document}
\title{Characterization of Surface Electromagnetic Waves and Scattering on Infrared Metamaterial Absorbers}

\author{Wen-Chen Chen$^1$, Machhindra Koirala$^1$, Xianliang Liu$^1$, Talmage Tyler$^2$, Kevin G. West$^3$,
Christopher M. Bingham$^1$, Tatiana Starr$^3$, Anthony F. Starr$^3$, Nan M. Jokerst$^2$ and Willie J. Padilla$^1$}
\email{Willie.Padilla@bc.edu} 

\affiliation{$^1$Department of Physics, Boston College, 140 Commonwealth Ave., Chestnut Hill, MA 02467, USA}
\affiliation{$^2$Department of Electrical and Computer Engineering, Duke University, Durham, NC 27708, USA}
\affiliation{$^3$SensorMetrix, Inc. San Diego, California, 92121, USA}

\begin{abstract}
We report, for the first time, a full experimental and computational investigation of all possible light matter interactions on the surface of an infrared metamaterial absorber (MMA). Two channels of energy dissipation - diffuse scattering and generation of surface electromagnetic waves - are quantified in terms of their impact on specular absorption. The diffuse scattering is found to play a negligible roll in the absorption process, at least for wavelengths greater than the periodicity of unit cell. In contrast, surface electromagnetic waves are found to be generated for transverse magnetic (TM) polarized light at the operational wavelength of the MMA, i.e. $\lambda_0$, and shorter wavelengths. Our computational results indicate that the highly lossy surface electromagnetic wave generated at $\lambda_0$  is responsible for the good angular dependence of absorption in TM polarization. Experimental results are supported by full wave three dimensional electromagnetic and eigenmode simulations.
\end{abstract}
\maketitle

Metamaterial absorbers (MMAs)~\cite{MPA_review} are a type of engineered material that obtain near-unity absorption of electromagnetic energy at a specified wavelength. The MMA operates by achieving an impedance ($Z(\omega)$) match to free space, while at the same time providing high loss - as determined by the imaginary portion of the index of refraction ($n_2(\omega)$). A unit cell of the MMA often consists of two metallic layers (typically one patterned and one continuous) with an insulating layer lying in-between. The unit cell is subwavelength and periodic in the plane. The patterned metallic layer is an electric ring resonator (ERR) \cite{ERR1,ERR2} and provides a Lorentz type resonant electric response ($\epsilon(\omega)$). The ERR interacts strongly with the underlying metallic layer and a resonant effective magnetic response ($\mu(\omega)$) can be obtained. The first realization of such a structure was experimentally demonstrated at microwave frequencies \cite{Landy_PRL08} and designs were quickly shown in the THz \cite{PATera1}, infrared \cite{XL1,Giesse2010} and optical regimes \cite{Atwater2011}. The thickness of the MMA is small compared to the free space wavelength of light thus yielding large absorption coefficients. The above mentioned salient features of the MMA suggest they may be used to replace conventional absorbers and to enable novel applications, such as frequency-selective detectors \cite{ds}, emitters \cite{XL2} and spatial light modulators \cite{XL1}.

\begin{figure}
\begin{center}
\includegraphics[width=3in,keepaspectratio=true]{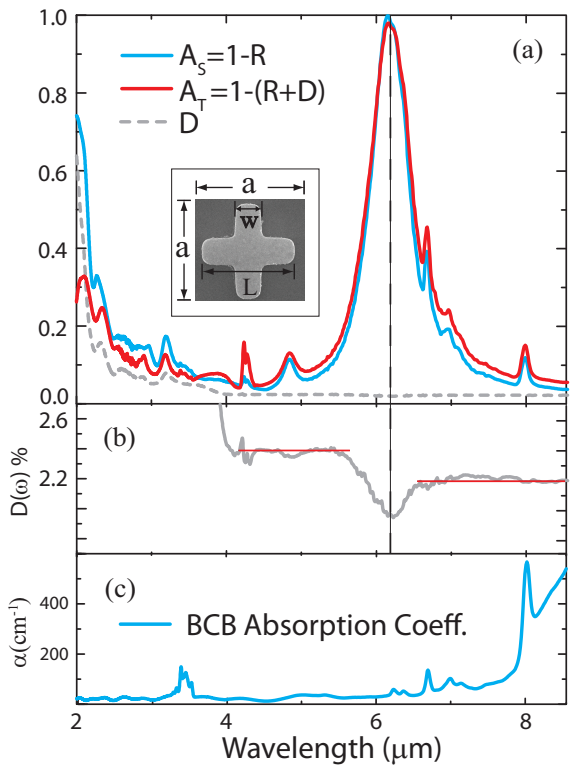}%
\caption{(a) Experimental diffuse scattering $D(\omega)$ (dashed curve), the specular absorbance $A_S$ (blue curve), and the total absorbance $A_T$ (red curve) of the IR MMA. The inset shows an SEM image of the fabricated sample with dimensions: $w$ = 500nm, $L$ = 1.95 $\mu$m, and $a$ = 3.2 $\mu$m. (b) An expanded view of $D(\omega)$ from 2-8.5$\mu$m detailing the global minimum at 6.13$\mu$m, as denoted by the vertical dashed line. (c) Experimental determined absorption coefficient of BCB.}
\label{fig1}%
\end{center}
\end{figure}

Metamaterial absorbers operating at infrared and optical wavelengths utilize a simpler ERR geometry compared to designs demonstrated at THz and lower frequencies \cite{XL1,Giesse2010,Atwater2011}. This permits the MMA to be fabricated more precisely, with less errors, thus producing higher yield and hence greater performance. The consequence, however, is that the unit cell size ($a$) is not as sub-wavelength as geometries utilized at lower frequencies. Experimental results have shown that the resonant wavelength to unit cell size ($\lambda_0/a$) is typically around 3 or 4 at infrared wavelengths. Although IR MMAs are still able to obtain high absorptivity, with values of 98\% shown \cite{XL1}, it is unknown when the scattering ($D(\omega)$) becomes non-negligible as the wavelength and unit cell size become close in dimension. Incident energy may further generate surface electromagnetic waves ($S(\omega)$), which may then be dissipated or re-radiated \cite{SEW1,SEW2,SEW3,SEW4}. Thus in a normal specular characterization, both diffuse scattering and excitation of SEWs may appear as absorption. To date no investigation has accounted for all possible light-matter interaction that may occur upon the surface of a MMA. Here we experimentally and computationally investigate the role of $D(\omega)$ and $S(\omega)$ to absorption in infrared MMAs. We demonstrate that scattering is small and not relevant in the absorption process, but that highly lossy surface electromagnetic waves are generated upon the MMA in TM polarization.

\begin{figure}
[ptb]
\begin{center}
\includegraphics[width=3.5in,keepaspectratio=true]{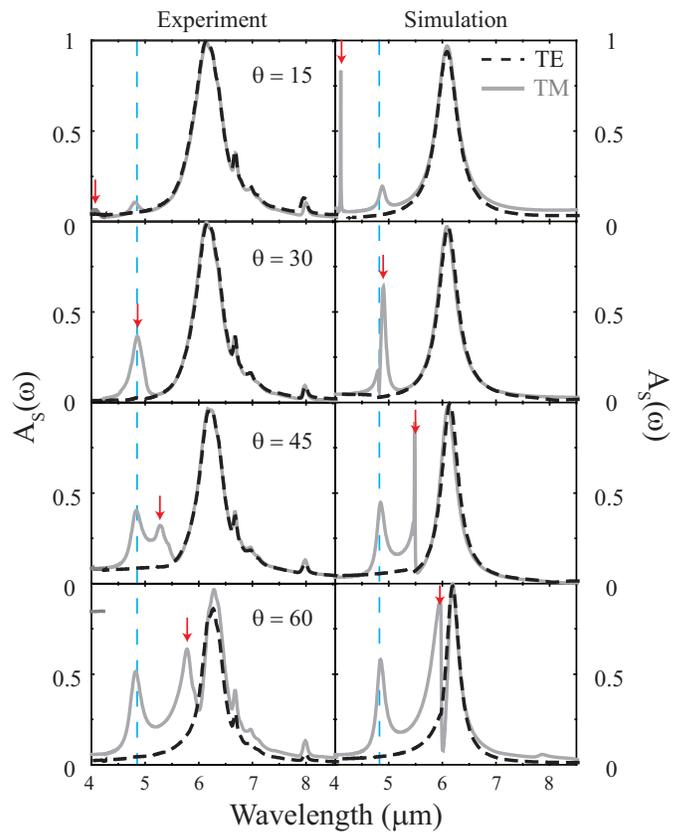}%
\caption{Experimental (left panels) and simulated (right panels) frequency dependent specular absorption for various incident angles ($\theta$) in both TM (grey solid curves) and TE (black dash curves) polarizations.}
\label{fig2}%
\end{center}
\end{figure}

The infrared metamaterial absorber sample was fabricated using thin-film processing combined with photolithographic patterning. The first step was the electron-beam vacuum deposition of a 3 nm Cr adhesion layer followed by a 100 nm thick gold ground plane. The dielectric spacer layer of the MMA is a 195 nm thick film of benzocyclobutene (BCB) which was spin coated onto the ground plane and thermally cured under vacuum. Next, deep UV lithography is used to pattern the metamaterial elements in the UV lithography resist, and a 3 nm Cr adhesion layer followed by 75 nm thick gold layer are then deposited onto the nanopatterned sample.  After performing lift-off of the Cr/Au, the remaining Cr/Au forms the MMA elements. A scanning electron microscope (SEM) image of a MMA unit cell is shown as the inset to Fig. 1(a). The patterned MM cross has the following dimensions: linewidth w = 500 nm, and side length L = 1.95 $\mu$m, with a unit cell size a = 3.2 $\mu$m. The fabricated samples have a relatively large surface area, with a total lateral size of $2.54 \times 2.54$ cm$^2$, in order to enable the accurate characterization of the MMAs at high incident angles.


We characterize the specular absorbance, defined as $A_S(\omega)=1-R(\omega)$, where $R(\omega)$ is the reflectance, shown as the blue curve in Fig. \ref{fig1}. (The transmittance $T(\omega)$ was zero in all measurements due to the thick continuous gold ground plane.) The reflectance (unpolarized) is collected at near normal incidence (21 degrees) and is normalized by the reflectivity of a gold mirror. We find that the IR MMA obtains a prominent peak with a value of $A_S$= 99.8\% at 6.14 $\mu$m. Outside of this main absorptive feature $A_S$ is relatively low and there are small local maxima observed at 6.67, 6.96, 7.10, 7.99, and 9.57 $\mu$m due to characteristic absorptions in the dielectric spacer layer of BCB \cite{uv_cure}, see Fig. \ref{fig1}(c) for detail.

In order to quantify the contribution of diffuse scattering to the specular absorption shown in Fig. \ref{fig1}, we characterized the MMA using a sphere integrator \cite{sp_int1,sp_int2}. A gold coated Lambertian scatterer (infragold$^{\circledR}$)\cite{infragold} is used to obtain the normalized diffuse scattered light ($D$) and normalized diffuse plus specularly reflected light ($D+R$). Radiation is incident at 13 degrees and measurements are performed at atmospheric conditions. In Fig. \ref{fig1} (b) $D(\omega)$ is plotted as the dashed grey curve and is low and relatively featureless from 4-8.5 $\mu$m with values around 2.2-2.3\% \cite{xray}. As we proceed to wavelengths shorter than 4 $\mu$m, $D$ begins to increase and is 64.0\% at 2.0$\mu$m. Interestingly the increase in $D$ is fairly rapid and continuous for wavelength shorter than 4$\mu$m, except for some notable features occurring at 2.32, 2.60, 2.89 and 3.20 $\mu$m. Having characterized the diffuse scattering we may plot the total absorbance, (solid red curve of Fig. \ref{fig1}(a)), defined as $A_T=1-(R(\omega)+D(\omega))$ over the range of 2-8.5 $\mu$m \cite{CO2_absorption}. It should be stressed that the two distinct measurements, $A_S$ and $A_T$, are in good agreement over the range from 4-8.5 $\mu$m, and only largely deviate from each other where the diffuse scattering becomes significant. Figure \ref{fig1}(b) shows a detailed view of $D(\omega)$ and it is found that a global minimum ($\sim 1.9\%$) in the diffuse scattering occurs at the exact position of the primary metamaterial absorption, denoted by the dashed vertical line.

Having verified the negligible impact of diffuse scattering on absorption we next investigate the possible existence of surface waves on MMAs and clarify their impact on the electromagnetic properties. We carry out a series of off-normal angle dependent reflectance measurements in both transverse electric (TE) and transverse magnetic (TM) polarizations. Experimental results for TE (dashed black curves) and TM (solid grey curves) at incident angles of 15, 30, 45, and 60 degrees are plotted as $A_S$ and shown in the left panels of Fig. \ref{fig2}. For both polarizations the spectral characteristics of $A_S$ are similar for near normal incidence (15 degrees) -- a main absorptive feature due to the MMA is observed near 6$\mu$m and other identifiable BCB absorptive signatures are found. Proceeding to larger incident angles we observe that, for the TE polarized case, the amplitude of the main peak drops noticeably from 90\% at 15 degrees to 75\% at 60 degrees. However, for TM polarization the peak in $A_S$ remains high for all angles investigated, but notably two distinct signatures are revealed for larger incidence angles. One of these absorptive features -- which we term here Mode A -- is independent of incident angle and occurs at a wavelength of 4.83$\mu$m. The other feature -- termed Mode B -- is observed to shift to longer wavelengths as angle is increased (as noted by the red arrows in Fig. \ref{fig2}).

In order to gain insight into the various absorptive features observed in the experimental results presented above, we perform 3D full wave electromagnetic field simulations (CST Microwave Studio 2012). The MMA was simulated with dimensions identical to that of the fabricated sample. We use a Drude model for all metallic components \cite{Ordal}, and a frequency independent dielectric constant of $2.06+i0.12$ for BCB. Unit cell boundary conditions are assigned which enable us to computationally investigate the angular and polarization dependent absorption. Simulations provide the complex scattering parameters and we calculate the specular absorption as $A_S=1-R=1-|S_{11}|^2$. The right panels of Fig. \ref{fig2} show $A_S$ for various incident angles and good agreement with experimental results is evident - notably both Modes A and B are also observed.

\begin{figure}
\begin{center}
\includegraphics[width=3.25in,keepaspectratio=true]{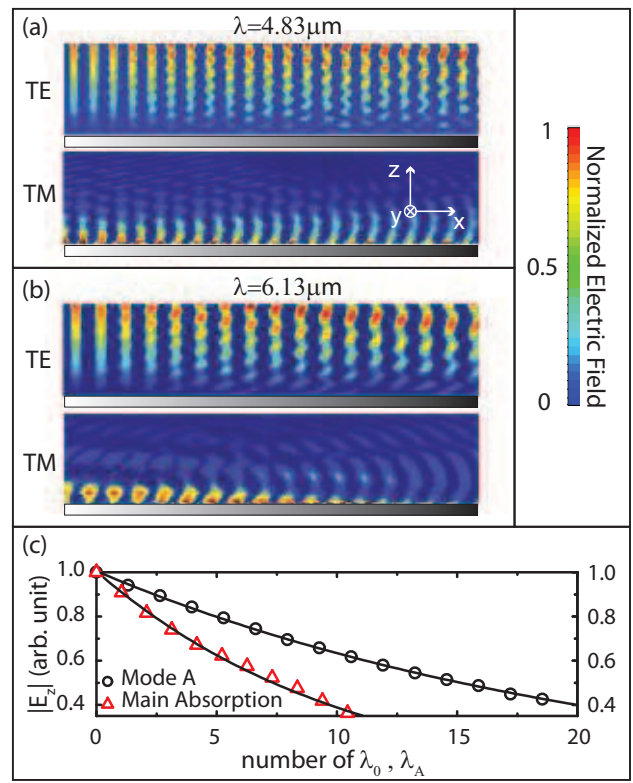}%
\caption{Simulated electric field along the surface of the metamaterial absorber at (a) 4.83 $\mu$m and (b) 6.14 $\mu$m for both TE, and TM polarizations. (c) $|E_z|$ as a function of position in the x-direction at 4.83$\mu$m (open triangles) and 6.14 $\mu$m (open circles). The horizontal axis is in units of the free space wavelengths of $\lambda_0$ and $\lambda_A$, where $\lambda_A$ denotes free space wavelength of Mode A.}
\label{fig3}%
\end{center}
\end{figure}

We next discuss the implications of the experimental and computational results and focus first on Mode B. We note that Mode B only appears in the TM polarization and its peak redshifts as a function of incident angle. It is well-known that periodic structures can support spoof surface-plasmon-polaritons (SPPs) for TM polarization incident waves \cite{Pendry_sicence2004}. Excitation of SPPs requires the momentum of incident light, $k_i$, to match that of the SPP \cite{SEW2,SpoofSPP_sicence2005}, i.e. $k_{spp}=k_i + k_G$, where $k_{spp}$ and $k_G$ are k-vectors of SPP and the Bloch wave (reciprocal lattice vector), respectively. By re-writing $k_i$ in terms of free space light $k_0$ as $k_i = k_0 sin\theta$, a strong angular dependence of such SPPs is evident. If Mode B is indeed a SPP, then its peak value should be a function of sin($\theta$). We perform a linear fit to our experimental data, (not shown) and find that our y-intercept, $k_G=2\pi / a$, equals a value of 1.71$\times 10^6$ (m$^{-1}$). A linear fit to the curve reveals a value of $a = 3.65$ $\mu$m, which agrees qualitatively with the lattice spacing (3.2$\mu$m) of the fabricated samples. Thus, we attribute the angular dependent feature, (Mode B), as a Bloch wave or spoof SPP, related to the periodicity of the metamaterial.

The other absorptive feature, (Mode A), however is independent of incident angle, always occurring near 4.83$\mu$m. Notably, Mode A seems to increase in strength as a function of increasing incident angle \cite{interference}.  Thus we computationally investigate its properties with light coming in at an incident angle of 90 degrees. A computational geometry consisting of 32 MMA unit cells is simulated and has a total length of 102.4$\mu$m, (see Fig. \ref{fig3}). Along the z direction, an open boundary is assigned to the top of the bounding box while a perfect electric boundary is assigned to the bottom (the metal ground plane). By setting either a perfect electric or perfect magnetic boundary condition in the y direction, incident light along the x direction can be approximated as TE or TM polarization.

The simulated transmission (not shown) shows two noticeable dips in TM polarization but, notably, is featureless for TE polarization, i.e. high transmission over the range investigated. Transverse magnetic polarized transmission minima occur at 6.17$\mu$m and 4.80$\mu$m, which can be correlated to the primary metamaterial absorption and Mode A in our measurements, respectively. In Fig. \ref{fig3} (a) and (b) we plot the electric field at 6.17$\mu$m and 4.80$\mu$m for both TM and TE polarizations. We first discuss $\lambda$=4.80$\mu$m, as plotted in Fig. \ref{fig3}(a), which shows a TE polarized wave which does not couple to the surface. However, drastically different behavior is observed for the TM polarized wave. We find that the TM incident wave is coupled to the surface of the metamaterial and both a decaying surface wave and a free space component are observed. The phase advance of the surface wave, compared to that of the free space wave, is different causing interference -- a signature of light coupling with surface plasmons. Although quantitatively the TE and TM polarized waves show disparate behavior, our simulations further allow us to extract quantitative information by approximating both the wave number and loss of the SEW. For example, in Fig. \ref{fig3} (a) the wavelength of the TM surface waves $\lambda_{A}$, may be determined by tracing the extrema of the electric field along the air/metamaterial interface. We find that the SEW yields a wavelength of about 4.72$\mu$m while the free space wavelength is 4.83$\mu$m. Thus this mode is a type of slow waves with a k-vector 1.02 times larger than that of free space light. In addition, the propagation length ($L_{spp}$) can be calculated by extracting the z component of the electric field as a function of position (x direction), see open black circles in Fig. \ref{fig3}(c). From a decaying exponential fit to the plot we find that $L_{spp}$ is about 102 $\mu$m indicating that the SEW travels over an appreciable distance of 25$\times$ its free space wavelength of 4.83$\mu$m. Similar behavior is observed for TE and TM polarized waves at $\lambda=6.13\mu$m, i.e. a SEW in TM polarization. We thus apply the same analysis in order to study the electric field surface profiles, see Fig. \ref{fig3}(b). One noticeable difference is the TM polarized SPP at $6.13\mu$m has a propagation length of 64.5 $\mu$m, which is only about half of that found for Mode A, shown in Fig. \ref{fig3}(c).

\begin{figure}
\begin{center}
\includegraphics[width=3.25in,keepaspectratio=true]{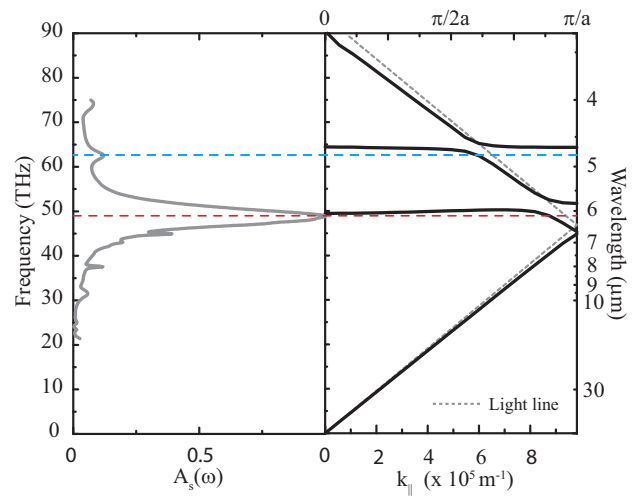}%
\caption{Experimental frequency dependent absorption (left panel) and simulated dispersion relation in the $k_{||}$ direction (right panel) for the metamaterial absorber. The two horizontal dash curves indicate the peak positions of mode A (blue) and the primary metamaterial absorptive resonance (red).}
\label{fig4}%
\end{center}
\end{figure}

We may gain further insight into Modes A and B by performing eigenmode simulations which allow us to calculate the band structure for the k-vector parallel ($k_{\|}$) to the MMA surface -- see  Fig. \ref{fig4}. An acoustic-like branch beginning from zero frequency is found with the slope equal to that of the free space light line. As can be seen, a band gap occurs before which another linearly dispersing mode is observed. In addition, two plasmon-like modes at 6 and 4.7$\mu$m can be observed. Although our eigenmode solver ignores losses, it should be stressed that these SPP-like modes occur at frequencies very close to the experimentally determined values of the primary absorption peak and mode A (red and blue dash curves of Fig. \ref{fig4}).

We computationally and experimentally investigated scattering and generation of surface electromagnetic waves on infrared metamaterial absorbers. Scattering was found to be negligible near the primary metamaterial absorption $\lambda_0$, and only became appreciable at shorter wavelengths. Two surface modes were found to exist on metamaterial absorbers, when incident light is TM polarized, one due to the periodicity of the unit cell and the other due to the effective optical constants of the MMA. The latter occurs at $\lambda_0$,  is highly lossy and is responsible for the good angular dependence of absorption in TM polarization.

We thank Dr. Weitao Dai for useful discussions. This research was funded by the Office of Naval Research under U.S. Navy contact No. N00014-10-C-0437, the NSF under Contract No. ECCS-1002340, and the Department of Energy under contract number DE-SC0005240.


\begin{thebibliography}{30}
\bibitem{interference} At 30 degress incidence, both Mode A and Mode B occur at the same wavelenght, thus increasing the apparent oscillator strength of Mode A.
\end{thebibliography}
\end{document}